\documentclass[12pt]{article}\usepackage[hyperfootnotes=false]{hyperref}
\usepackage{epsfig}
\usepackage{amsmath}
\usepackage{amssymb}
\usepackage{graphicx}
\newlength{\abstractwidth}
\renewcommand{\thefootnote}{\fnsymbol{footnote}}
\renewcommand{\thanks}[1]{\footnote{#1}} 
\newcommand{\starttext}{
\setcounter{footnote}{0}
\renewcommand{\thefootnote}{\arabic{footnote}}}
\renewcommand{\theequation}{\thesection.\arabic{equation}}
\newcommand{\be}{\begin{equation}}
\newcommand{\bea}{\begin{eqnarray}}
\newcommand{\eea}{\end{eqnarray}}
\newcommand{\beq}{\begin{equation}}
\newcommand{\ee}{\end{equation}}
\newcommand{\eeq}{\end{equation}}

\def\ba{\begin{eqnarray}}
\def\ea{\end{eqnarray}}

\def\12{{1 \over 2}}

\def\ra{\rangle}

\def\simleq{\; \raise0.3ex\hbox{$<$\kern-0.75em
\raise-1.1ex\hbox{$\sim$}}\; }
\def\simgeq{\; \raise0.3ex\hbox{$>$\kern-0.75em
\raise-1.1ex\hbox{$\sim$}}\; }

\def\O2{\Omega_2}

\def\bi{\begin{itemize}}
\def\ei{\end{itemize}}

\def\sc{\setcounter{equation}{0}}

\def\W{$\Omega$}
\def\W'{$\Omega$}

\def\V{\Omega}
\def\V'{\Omega}

\def\O{${\cal{O}}$}

\def\c{{\cal{C}}}

\def\bn{\bigskip \noindent}

\begin{document}
\renewcommand{\theequation}{\thesection.\arabic{equation}}
\begin{titlepage}
\rightline{}
\bigskip
\bigskip\bigskip\bigskip\bigskip
\bigskip
\centerline{\Large \bf { Butterflies on the Stretched Horizon}}
\bigskip
\begin{center}
\bf Leonard Susskind \rm

\bigskip

\bigskip

Stanford Institute for Theoretical Physics and Department of Physics, \\
Stanford University,
Stanford, CA 94305-4060, USA \\
\bigskip
\bigskip
\vspace{2cm}
\end{center}
\bigskip\bigskip
\bigskip\bigskip
\begin{abstract}
In this paper I return to the question of what kind of perturbations on Alice's side
of an Einstein-Rosen bridge can send messages to Bob as he enters the horizon at
the other end. By definition ``easy" operators do not activate messages and ``hard"
operators do, but there are no clear criteria to identify the difference between easy
and hard. In this paper I argue that the difference is related to the time evolution of
a certain measure of computational complexity, associated with the stretched horizon
of Alice's black hole.

The arguments suggest that the AMPSS commutator argument is more connected
with butterflies
 than with firewalls.

\medskip
\noindent
\end{abstract}
\end{titlepage}
\starttext \baselineskip=17.63pt \setcounter{footnote}{0}
\tableofcontents

\sc
\section{Complexity and Geometry}
Computational complexity \cite{watrous} does not play much of a role in gravitational physics  at present (see however \cite{Harlow:2013tf}), but  that may  change. I will argue that  complexity plays a key role in both the deep inner structure of the stretched horizon and the extreme long-time behavior of black holes.  It determines what actions at one end of an Einstein-Rosen bridge can send signals to the other end\footnote{The  term  ``end of the Einstein-Rosen bridge" refers to points just behind the horizon of one of the black holes. Nothing in this paper should be interpreted as violating locality by traversing wormholes. }.

 The framework for this paper is  gauge-gravity duality \cite{Maldacena:1997re}   involving a bulk gravitational theory in ADS and a dual gauge theory. In order to count degrees of freedom \cite{Susskind:1998dq}  the scale invariance of the field theory must be broken by introducing a regulator. The details are not important but for conceptual clarity the language of a lattice theory will be used.
 Each lattice point represents $N^2$ degrees of freedom and is therefore as rich as an entire bulk region of size $l_{ads},$ i.e., the radius of curvature of ADS.

Take a local unitary operator $E$ such as the limit of a small Wilson loop: if one acts with $E$  a localized disturbance is created at one site. That's about as simple a disturbance as one can consider, in that it only depends on the degrees of freedom at one lattice site. Now let us evolve $E$ in time.

$$E \to U(t) E U^{\dag}(t)$$

\bigskip
\noindent
The resulting operator is not as simple as the original $E;$ it involves all the operators in a growing region of space. One can say that the complexity of the operator, or the disturbance it creates, is increasing. One way of describing the complexity of the disturbance  is to ask how many elementary computations are needed to implement the operator.
The growing complexity is connected, through the UV/IR connection with the motion of a signal from the boundary into the bulk of ADS. The tendency for complexity to increase is related both to the tendency to spread into the IR, and to the tendency to fall toward the horizon of ADS.

Go back to the simple operator $E$ and run it backward in time. It will not
 get any simpler; in fact it will get more complex. The simple disturbance is at a local minimum of complexity---the complexity grows in either time direction. Corresponding to that, a particle coming up from the bulk of ADS will reach a certain height and fall back. In what follows I will advocate a relations between complexity and radial distance, and  between decreasing/increasing complexity and rising/falling in the gravitational field of ADS. But my real motivation is understanding the ER=EPR connection \cite{Maldacena:2013xja} and the ability of Alice to send messages to Bob.

 According to  \cite{Maldacena:2013xja} the only way to create a firewall behind Bob's horizon is to send it through the Einstein-Rosen bridge connecting Bob's black hole to its purification. The purification could be the black hole's Hawking radiation, a dust cloud, or another black hole. The complexity arguments in this paper indicate that sending even a simple message can be incredibly hard, comparable in difficulty to reversing the arrow of time for a large complex system.

\sc
\section{String Model}

In discussing  the microscopic degrees of freedom of a black hole we would like to have a specific model to work with. Generally all we can say about the degrees of freedom is that there are $$\frac{A}{4G\hbar}$$ of them and that they in some sense occupy a stretched horizon. Even in string theory, the detailed properties of the stretched horizon  are elusive---with one exception:
 A Black hole and a  highly excited string of the same entropy can be regarded as two ends of an adiabatic evolution, in which the string-coupling is gradually turned on. A crossover occurs at the value of the coupling at which the string length\footnote{By the string length I mean $\sqrt{\alpha'}$} $l_s$ equals the Schwarzschild radius of the black hole. At the crossover the system can be analyzed as a string, or as a black hole, with results that often agree surprisingly well \cite{Susskind:1993ws}\cite{Sen:1994eb}\cite{Russo:1994ev}\cite{Peet:1995pe}\cite{Halyo:1996xe}\cite{Horowitz:1996nw}.

 A highly excited string can be visualized as a long chain of links \cite{Karliner:1988hd} \cite{Mezhlumian:1994pe}, each of length $l_s.$ The length, entropy, and energy of the string are all proportional to the number of links which we take to be $N^2$. We may also think of the orientation of the links as qubits \cite{Klebanov:1988ba}.  The evolution of such a system can be regarded as a quantum computation.

The typical highly excited state of a single string consists of a random walk of length $L,$  mass $M,$ and entropy $S.$ These properties are related:

\be
L   = l_s S
\label{L = ls S}
\ee

\be
M=  \frac{S}{l_s}
\label{M = S/ls}
\ee

 \bigskip
 \noindent
 The string length $l_s$ is related to the Planck length by

 \be
 l_p = g^2 l_s
 \label{l_p = g^2 l_s}
 \ee

 \bigskip
 \noindent
 and the crossover or correspondence point is defined by setting the Schwarzschild radius to the string length,

 \be
 R_s = l_s
 \label{Rs = ls}
 \ee

  \bigskip
 \noindent
 or

  \be
M {l_p}^2 = l_s
 \label{crossover}
 \ee

  \bigskip
 \noindent
 Notice that if we combine \ref{crossover} with \ref{M = S/ls} we get the standard black hole relation

 \be
 S = M^2 l_p^2,
 \label{S(M)}
 \ee

  \bigskip
 \noindent
and using \ref{Rs = ls}, the Bekenstein relation,

 \be
 S = \frac{R_s^2}{l_p^2}
  \label{Bekenstein}
 \ee

\begin{figure}[h!]
\begin{center}
\includegraphics[scale=.5]{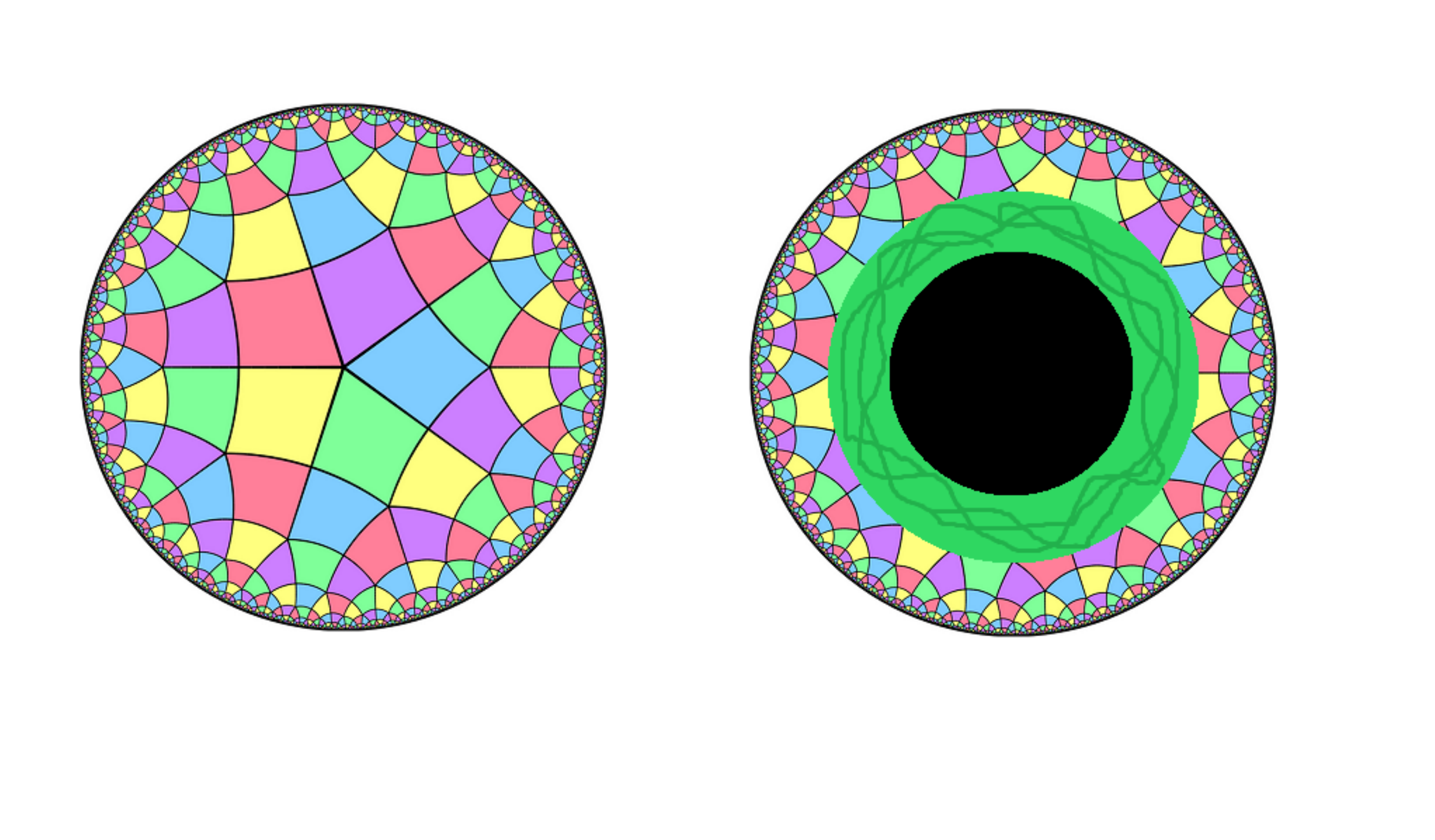}
\caption{ ADS and ADS with a unit black hole. The green region indicates the stretched horizon which extends to a distance $l_{ads}$ above the horizon.}
\label{f1}
\end{center}
\end{figure}

   \bigskip
 \noindent
 Now let's consider ADS black holes. The smallest stable ADS black hole is at the Hawking-Page transition \cite{Hawking:1982dh}. It has entropy of order $N^2,$ where $N$ is the rank of the dual gauge theory; Schwarzschild radius of order $l_{ads}$ ; and mass of order $N^2/l_{ads}.$   The outer edge of the stretched horizon is an ADS distance from the horizon of the black hole.  Figure \ref{f1} is a representation of such a unit black hole.

These ADS-size stable black holes are  units, out of which larger higher temperature black holes can be built.
 If the gauge theory is regulated near the thermal wavelength, then each lattice site is a unit black hole with entropy $N^2.$ A larger
ADS black hole qualitatively behaves like  a spatial lattice of such unit black holes.
 Therefore, from now on  we  work with the unit black holes at the Hawking-Page transition.

One other quantity must be specified, namely   the 't Hooft coupling \cite{'tHooft:1973jz}

$$\lambda =g^2N$$

   \bigskip
 \noindent
 we take to be of order  unity. This puts the theory right at the crossover point where black holes transition to single long strings \cite{Susskind:2013lpa}. The value of the string length and the ADS length are the same at this point,

\be
l_s = l_{ads}
\ee

  \bigskip
 \noindent
In this picture a unit black hole is a single long string of  $N^2$ ADS-length links. The entropy is stored in the configuration of these segments. Roughly speaking each pair of neighboring segments may be thought of as a qubit representing the way the string turns from segment to segment.

Another way to view the long-string model is to begin with a very coarse-grained  version of the dual gauge theory. Given that $\lambda \sim 1,$ the useful gauge invariant regulated description is  Hamiltonian lattice gauge theory \cite{Kogut:1974ag}.
From the UV/IR connection, one expects that the degrees of freedom of a unit black hole are captured  by a large $N$ lattice gauge theory containing just a small number of lattice sites; for example a lattice gauge theory on a single cubic cell.
At large $N$ the theory has a thermal de-confinement transition, similar to the Hawking page transition, at which the thermal entropy is dominated by states of a single long  chromo-electric flux string.

The stretched horizon of a string-like black hole has a thickness of order $l_s.$
 In the present case the string scale  is the ADS radius, so that the stretched horizon extends out to a distance $l_{ads};$ in other words to the outer boundary of the green region in figure \ref{f1}.

The idealized strings of string theory are integrable and have exact degeneracies which we don't expect to find in black holes. The energy spectrum of a black hole is chaotic and consists of non-degenerate energy levels. The average spacing is $e^{-S}$ but with statistics typical of quantum-chaotic systems.
Since chaos will play a key role in what follows we must assume that the thick gauge-theory strings that model a unit ADS black hole are not integrable. We will assume that they have the properties of chaotic systems\footnote{In the large $N$ limit string theory on $ADS_5 \times S_5$ is integrable if one holds the excitation level fixed. But in the limit we are working in, where  the energy of the string is allowed to increase like $N^2,$ string interactions are never negligible and there is no reason for such long strings to be integrable.}. For example, they thermalize and scramble. Local excitations evolve by diffusing over the length of the string.

The diffusive properties allow us to make a correspondence similar to the usual UV/IR correspondence.
Let us consider the long string to be in thermal equilibrium and consider a local perturbation. For example, acting with a single plaquette in the lattice theory creates a perturbation on a small number of neighboring sites along the string. Such a localized perturbation can be called ultraviolet, but with respect to the string, not the dual field theory.

\begin{figure}[h!]
\begin{center}
\includegraphics[scale=.3]{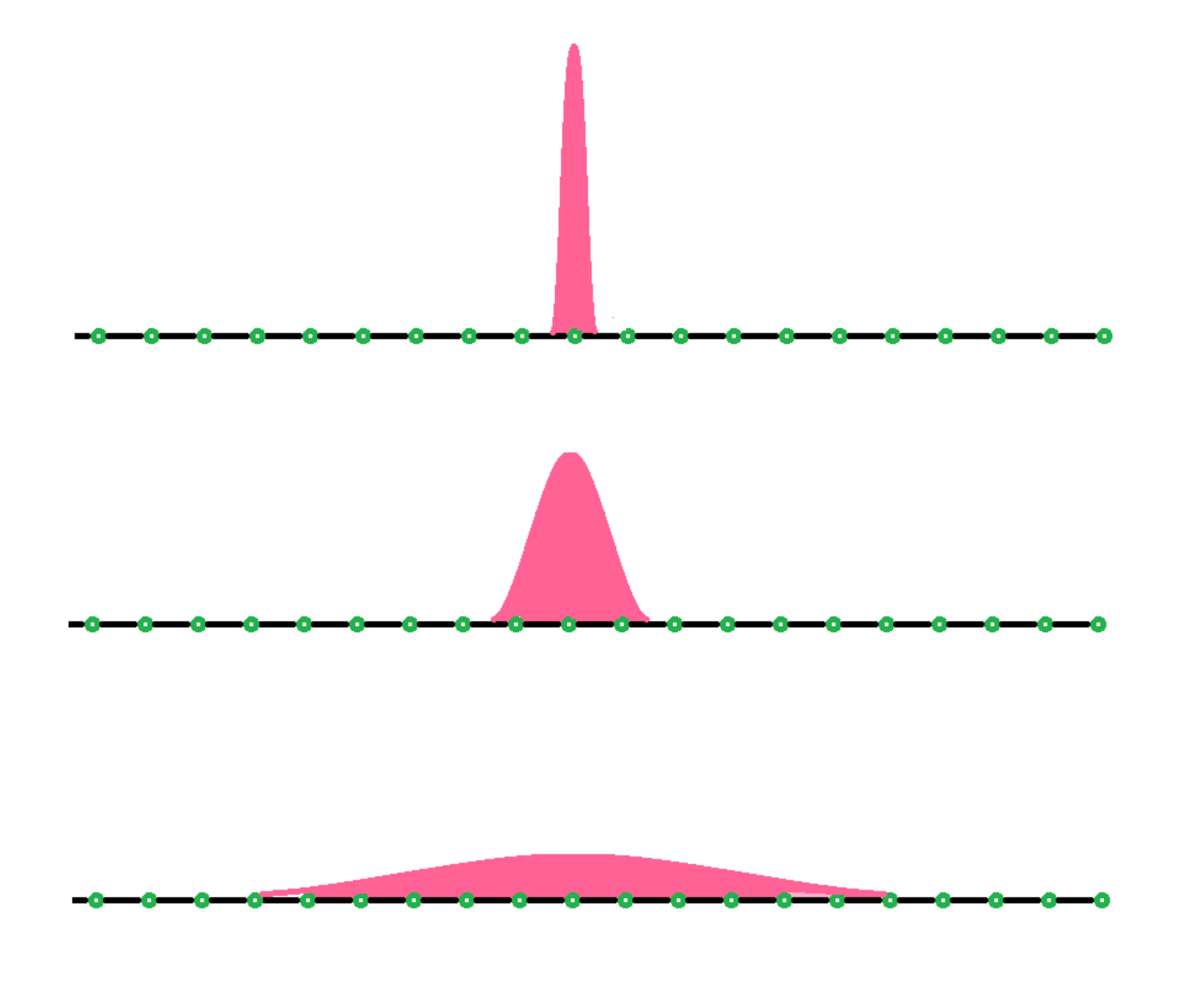}
\caption{ A localized disturbance on a string diffuses.}
\label{f2}
\end{center}
\end{figure}

In time, diffusion will spread the UV perturbation into the long wavelength (IR)  modes of the string as indicated in figure \ref{f2}.
The same thing is true of the gauge theory itself; in thermal equilibrium a localized excitation will diffuse outward. In both cases the outward diffusion can be identified with the tendency to fall toward the horizon. These twin tendencies---diffusing and falling toward the horizon---are also connected with the entropic description of gravity \cite{Verlinde:2010hp}  in the vicinity of  a horizon.

Black holes are fast scramblers that scramble quantum information very efficiently---in a  scrambling time $t_{\ast}$ given by

\be
t_{\ast} = l_{ads}  \ \log{\left(\frac{l_{ads}}{l_p}\right)}.
\label{tstar}
\ee

   \bigskip
 \noindent
  If we replace the usual dimensional time variable by the dimensionless hyperbolic angle $\omega$ then the scrambling time is

  \be
\omega_{\ast} =  \ \log{S}.
\label{ostar}
\ee

   \bigskip
 \noindent
   We  will assume that that the dynamical evolution of the string is governed by a Hamiltonian

\be
U(t) = e^{iHt}.
\ee

   \bigskip
 \noindent
which scrambles according to \ref{tstar}.
The ordinary lattice gauge Hamiltonian has properties which suggest that it is a fast scrambler \cite{Page:1993wv}\cite{Hayden:2007cs}\cite{Sekino:2008he}\cite{Lashkari:2011yi}\cite{Heemskerk:2012mn}.

\sc
\section{Complexity}

The complexity of a particular operation can be measured by the minimum number of elementary steps that it takes to carry out the operation. Suppose we begin with a  known state of the long string and apply a recognizable local perturbation to a couple of neighboring sites. We can do so by acting with a unitary operator $E$ that acts on a single qubit in the chain\footnote{In references \cite{Maldacena:2013xja} and \cite{Susskind:2013lpa} the notation $E$ referred to easy in the sense of easy to measure. The meaning is the same here. The reason that $E$ is easy to access is that it only involves a small number of ``computational qubits"}.
By recognizable I mean that it sticks out above the thermal fluctuations.
But now let the system evolve for a time. The perturbation will diffuse and become harder to recognize. Precursor operators are a useful way to think about the evolution of disturbances \cite{Polchinski:1999yd}\cite{Giddings:2001pt}\cite{Freivogel:2002ex} operator\footnote{As in \cite{Susskind:2013lpa} I will use precursor to indicate operators \ref{precursor} for either positive or negative $t.$ Later I will distinguish future and past precursors.}. A precursor of $E$ is defined in terms of the time development operator $U$ by,

\be
E_P= U(t) E U^{\dag}(t)
\label{precursor}
\ee

   \bigskip
 \noindent
will be spread over many links or qubits, the number depending on the time $t.$ By the scrambling time it will be spread over the entire string.

If we know $E$ and the evolution operator we can always undo the effects of $E_P$
by running the system through a quantum computer.  The relevant question is not how hard it is to know  $E_P,$ but given $E_P$ how hard it is to actually implement  it on a quantum computer. The number of steps that it take to implement $E_P$ is what I mean by its complexity.

We obviously  need a rule about what operations (of the quantum computer ) are allowed. If the quantum computer can apply any unitary operator in a single step, then nothing is complex. The standard rule is to assume that computations are built out of elementary units (gates), each of which involves a small number of qubits. A common example is to allow the gates to be one and two qubit gates. A two qubit gate is a unitary operator acting on two qubits.
The complexity $C$ of a unitary operator  acting on a set of $n$ qubits is the minimum number of gates that are needed to implement it.
\bigskip

\bn
There are two important facts about complexity that we will return to. The first is that scrambling can be accomplished with a rather small number of gates. For a system of $N^2$ qubits\footnote{In this context one usually defines $N$ to be the number of qubits. Here $N$ refers to the rank of the dual gauge group, so the number of qubits is $N^2.$} scrambling can be accomplished with $N^2 \log{N^2}$ gates \cite{Hayden:2007cs}. A scrambled system may look extremely random, but its complexity is very modest. The scrambling complexity is

\be
C_{scr} = N^2\log N^2.
\label{Cscram}
\ee

\bn
By the scrambling time the initially affected qubit will have interacted with every other qubit. However \ref{Cscram} is by no means the maximum complexity. Between scrambling and Haar-randomness \cite{Hayden:2007cs} there lies a vast range of complexity that involves extremely subtle correlations.

The second fact is that there is a maximum complexity $C_{max}$ far beyond
\ref{Cscram}.  Any unitary operator can be implemented with an exponential number of gates \cite{Harlow:2013tf}\cite{Kitaev}\cite{preskill}\cite{Nielsen}. The precise coefficient in the exponent is not important and I will ignore it. Thus we may say that the maximum complexity  is,

\be
C_{max} = e^{N^2}.
\label{Cmax}
\ee

   \bigskip
 \noindent
The gates can be considered to act in succession---one after another---but this is not the best way to model the evolution of systems of many degrees of freedom. In such systems, degrees of freedom can interact at the same time in small groups.
 To model this it is useful to think of the gates as acting in parallel. At any given time-step, we may allow $N^2/2$ gates to act simultaneously \cite{Hayden:2007cs}. Each qubit participates in a single two-qubit gate in each step. The natural time-step is the inverse temperature \cite{Sekino:2008he}.

 We will redefine the complexity of a unitary operator to be the number  of such parallel operations that are required to build the operator and call it $\c.$  With this definition scrambling can be accomplished with complexity,

\be
\c_{scr} = \log N^2 = \log S.
\label{cscr}
\ee

   \bigskip
 \noindent
Complexity like entropy has a tendency to increase with time.
Keeping in mind that the scrambling time satisfies \ref{ostar}, equation \ref{cscr} suggests the following interpretation. At every unit interval of $\omega$ one parallel operation occurs in which $N^2/2$ gates act simultaneously. Moreover, this dynamics is efficient in the sense that the  number of $\omega$-intervals represents the minimum number of such steps that are needed. Thus the complexity of the black hole evolution grows linearly with time,

\be
\c(t) = \omega = \frac{t}{l_{ads}}.
\label{c=omega}
\ee

   \bigskip
 \noindent
The growth of complexity will proceed until it reaches the maximum \ref{Cmax} at the saturation time $\omega = e^S.$ The saturation time happens to also be the classical recurrence time.

What happens to the complexity after this saturation time? The answer that it remains constant for a time of order the quantum recurrence time \cite{Bocchieri}. The quantum recurrence time is doubly exponential. Among other things it is the time for a recurrence in which $U$ becomes very close to its initial value $U=1.$ At this time the complexity will have recurred to its minimum value. Thus, like entropy, complexity can go either way; it sometimes increases and sometimes decreases. But if it is not near the maximum, it is overwhelmingly likely that it will increase in the next time interval---either to the future or to the past.

\sc
\section{Complexity and the Stretched Horizon}
\begin{figure}[h!]
\begin{center}
\includegraphics[scale=.3]{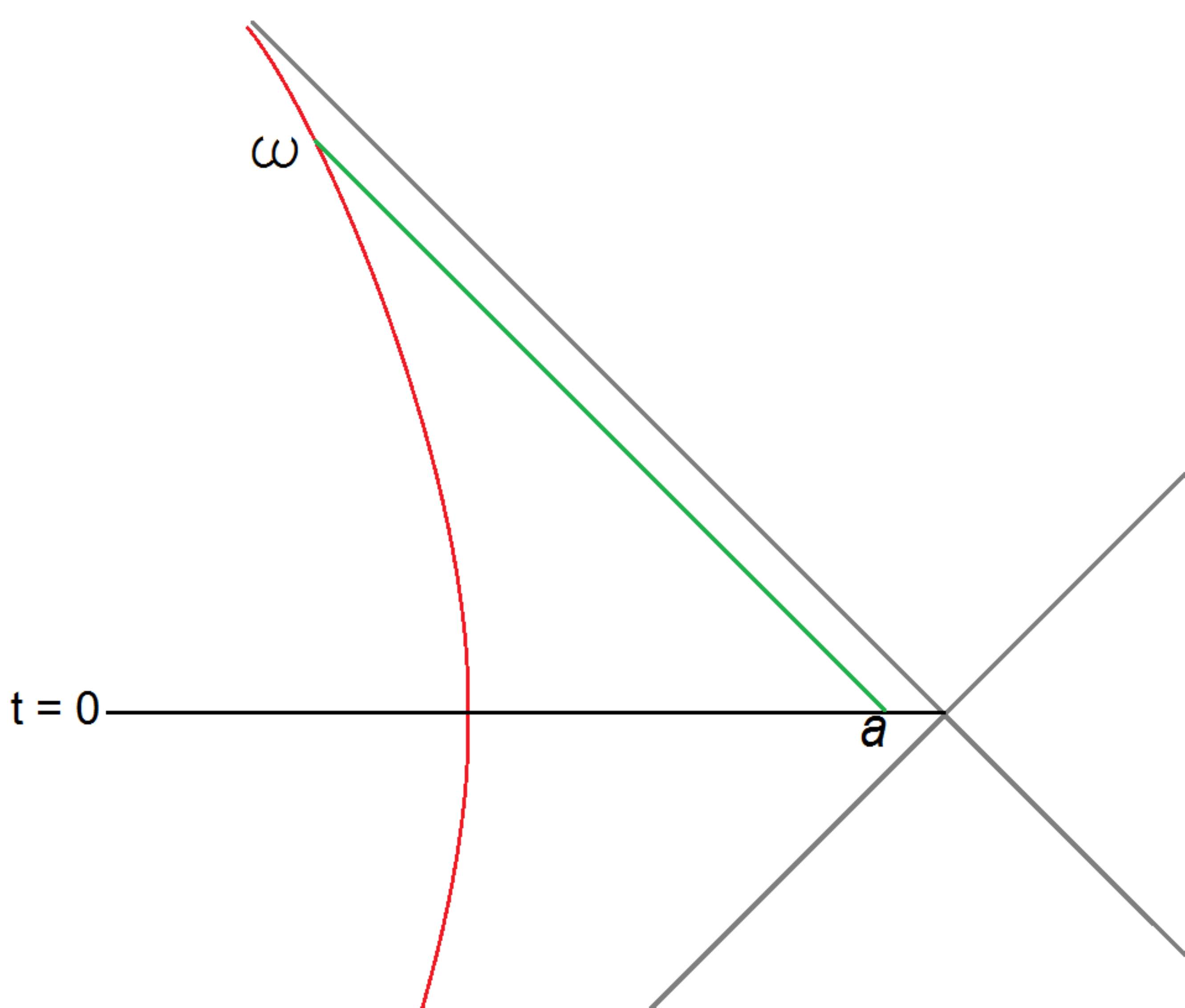}
\caption{ The red curve is the outer edge of the stretched horizon. An observer at time $\omega$ looks back to point $\bf{a}.$ This defines a mapping from $\omega$ to a layer at distance $l_a.$ }
\label{f3}
\end{center}
\end{figure}

I will propose a connection between complexity and the geometry of the stretched horizon  (SH).
The SH of an ADS black hole at 't Hooft $\lambda =1$ is rather thick, lying between the mathematical horizon and a surface about one ADS length above it. The connection with complexity involves assigning a complexity to every layer within the SH. The construction, shown in figure \ref{f3} begins with an observer stationed just outside the stretched horizon at a distance $ l_{ads}$ from the proper horizon. At time $\omega,$ the observer looks back to the surface $t=0$ and sees a point $\bf{a}. $ The point is a distance

\be
l(a) = e^{-\omega} l_{ads}
\label{layer}
\ee

   \bigskip
 \noindent
from the proper horizon. Thus we identify a time $\omega$ with a layer of the stretched horizon according to \ref{layer}.

Now using \ref{c=omega} we may also make an identification between complexity and $l(a),$

\be
l(a) = e^{-\c} l_{ads}
\label{layer2}
\ee

   \bigskip
 \noindent
The value of $l(a)$ for scrambling time is given by the Planck distance,

\be
l_{\ast} = l_p.
\label{plancklayer}
\ee

Beyond the scrambling time the identification becomes very unstable due to the butterfly effect, as we will see later in Sections 8 and 9. A formal continuation of \ref{layer2} to large times would identify the maximum complexity with a minimimum distance from the horizon:

\be
l_{\min}= e^{-e^S}
\ee

   \bigskip
 \noindent

\sc
\section{The Thermofield Double}

Our focus will be on the ``two-sided" ADS system consisting of two uncoupled conformal field theories, each with a gravity dual.
The thermofield double (TFD). The TFD represents the initial state of
 a pair of maximally entangled ADS black holes \cite{Israel:1976ur}  located on completely separate spaces, with no interaction between them \cite{Maldacena:2001kr}\cite{Hartman:2013qma}.  The fact that they share a common Einstein-Rosen bridge represents their maximal entanglement. The entangled TDF state had the form

\be
|TFD\ra = \sum_n e^{-\frac{\beta}{2} E_n}|n, n\ra
\ee

  \bigskip
 \noindent
where $|n, n\ra$ labels energy eigenvectors of the gauge theories and $\beta,$ the inverse temperature, is of order the ADS radius,

\be
\beta \approx l_{ads}
\ee

  \bigskip
 \noindent
 Modeling the TFD state in the string description of the black hole is straightforward. We consider two uncoupled strings, each of length $N^2$ in ADS units. The strings are put into a maximally entangled configuration. In the lattice version the configurations of the strings are discrete and one may write

 \be
 |TDF\ra = \sum_n |n, \bar{n}\ra
 \ee

   \bigskip
 \noindent
 where $n$ represents a closed configuration of $N^2$ links and $\bar{n}$ represents its mirror image. The sum is over all configurations of the links.

 In many respects the TFD state is the simplest maximally entangled state. It is analogous to a unscrambled product of Bell pairs. In \cite{Susskind:2013lpa} this was described by saying that the TFD has no vertical entanglement. The development of vertical entanglement as the TDF evolves was described by Hartman and Maldacena \cite{Hartman:2013qma}.

\begin{figure}[h!]
\begin{center}
\includegraphics[scale=.3]{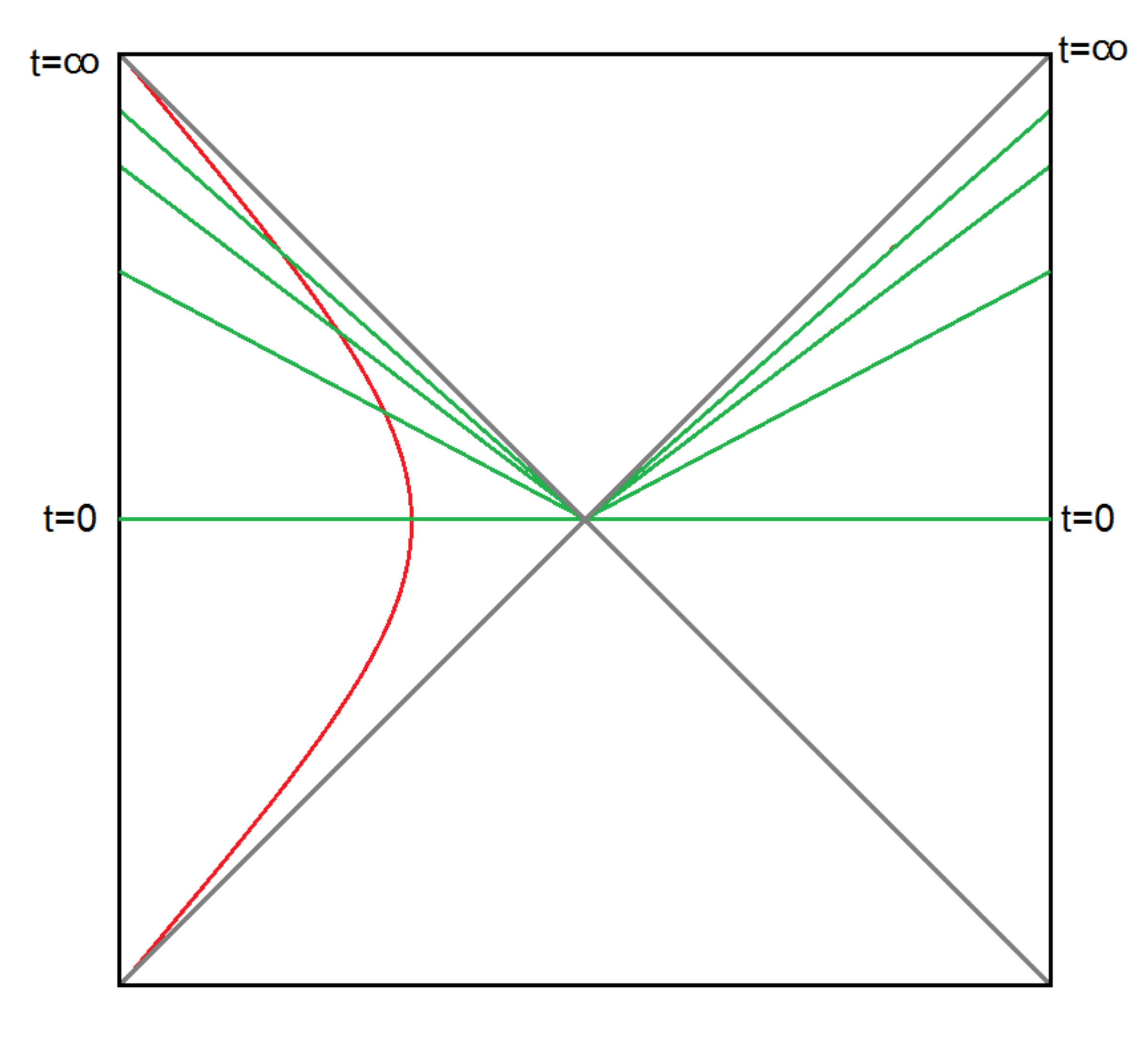}
\caption{The thermofield double. The red line is the outer boundary of the stretched horizon. The green lines are surfaces of constant time. }
\label{f4}
\end{center}
\end{figure}

\begin{figure}[h!]
\begin{center}
\includegraphics[scale=.3]{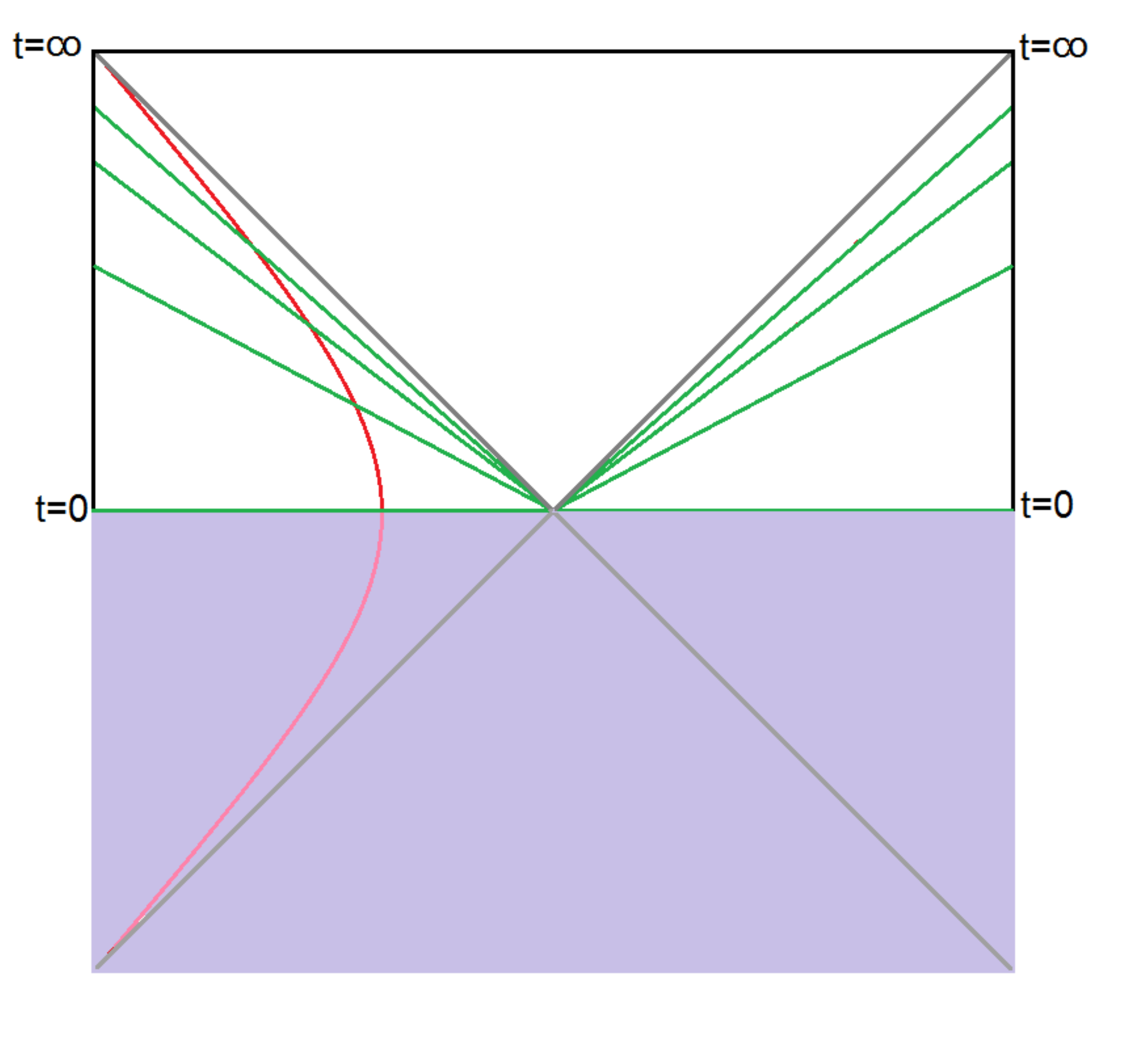}
\caption{The lower half of the diagram is fictitious. }
\label{f5}
\end{center}
\end{figure}

 The TFD state is identified with the eternal ADS black hole geometry represented by the Penrose diagram \ref{f4}.
The geometry represented by this Penrose diagram is eternal both to the future and the past. In what follows we will assume that only the future half of the diagram is physical. The past half, $t<0$ is a fiction representing the process that created the entangled system at $t=0.$ In figure \ref{f5} the diagram is redrawn with the lower half shaded out.

The reason for considering the lower half of the diagram to be un-physical is connected with the main theme of this paper, namely fine-tuning of initial conditions. There are initial conditions which are so fine-tuned that the increase of entropy seems to be reversed. In a gravitational context these reverse phenomena are connected with past singularities and are called white holes. After the white hole is created it is unstable to tiny effects, and therefore such initial conditions are generally deemed un-physical. Page has argued that past singularities should be censored for this reason \cite{Page:2013mqa}.

\subsection{Addition to Version 2}

\bf
This does not  mean that creating the initial TFD state is unreasonably hard. The difficulty is comparable to the following challenge:

 Starting at $t=0$ with two sets of $N^2$ qubits, all in the state $|0\ra,$ create a state consisting of a product of $N^2$ Bell pairs.  This can be accomplished using only $N^2$ gates. Since the gates act in parallel there is no issue of errors compounding. Therefore the construction of the TFD is not very difficult.

What it does mean is that attempting to set up the TFD by constructing its ancestor in the distant past is extremely difficult.
 First of all more gates are required and secondly, the system is extremely unstable during the lower half of the diagram. An example of this instability is the butterfly effect described by Shenker and Stanford \cite{Shenker:2013pqa}.

 This situation of an un-physical past is not uncommon, another case being  vacuum decay. The Minkowski continuation of the Coleman-DeLuccia instanton describes a bounce, but only the future half of the solution is physical.
The creation of the TFD state at $t=0$ may seem artificial, but it shares two  important features with the creation of an ordinary black hole by collapse:

1)  The preparation is not  difficult.
 
2)  The initial black hole is guaranteed to have a smooth horizon for some period after it has formed.
 
 \bigskip
 
 \bigskip

It is tempting to propose the following slogan:

\bigskip

 Black holes that start smooth, stay smooth\footnote{The possible exception to this rule (See Section 11) involves time scales of order the Poincare recurrence time. This would be relevant only for non-evaporating black holes. }.

\rm

\bigskip

\sc
\section{Messages From Alice to Bob}

The question raised in \cite{Susskind:2013lpa} is whether Alice, who has complete control over the  degrees of freedom of the left CFT system, can send a signal that Bob, on the right side, can receive. The message would travel through the Einstein-Rosen bridge and be received only after \rm Bob jumps through the horizon of the black hole on the right side. On the face of it the answer would seem to be no. Suppose Alice sends a message by creating a local perturbation of the left CFT. The earliest Alice can do this is $t=0$ since the entire system was not created until that time. If we assume that  Bob cannot jump into his black hole before $t=0$ then it looks like the best that Alice can do is to send a message that barely gets to Bob just as he hits the singularity. But as described in \cite{Maldacena:2013xja} and \cite{Susskind:2013lpa}, if Alice can apply non-local CFT operators she can reach into the bulk and send a message from a point that can reach Bob earlier.

Let's make Alice's task more difficult by postponing both her action, and   Bob's jump, to a later time $t.$  In her first attempt she acts by applying an easy operator $E,$ i.e., an operator that affects a only a few  adjacent links.
The easy operator is minimally non-local and not complex at all. Therefore it should be localized at a distance $\sim l_{ads}$ above the horizon. It is evident that acting with $E$  will not get a signal to Bob, but for later purposes it will be useful to explain why in the complexity language.

Alice acts at time $t$ with an easy operator $E$ but let's  agree to replace her actions by an equivalent past-precursor at $t=0.$ To indicate a past-precursor of a given operator I will add a superscript negative sign. Thus the past precursor of $E$ is defined by,

\be
E_P^- = U^{\dag}(t) E U(t).
\label{E'=UEU^*}
\ee

   \bigskip
 \noindent
To understand $E_P^-$ let us think of it as the backward evolution of $E$ for a time $t.$ If we evolve a very simple operator like $E$ for a time $t$ it will become more complex. It will diffuse over the string and involve a greater number of string segments. To put it another way the operator flows from the UV to the IR as we run backward. This would be  true whether we evolve forward in time or backward. Basically it is a manifestation of the second law. By the UV/IR connection the operator $E_P^-$ must be localized closer to the horizon than the operator $E.$
The same thing would be true if we use the complexity of $E_P^-$ to judge its location according to \ref{layer}.

If the complexity of $E_P^-$ increases as we run the time backward, then it decreases as we go forward from $t=0.$ This may be interpreted as saying that the perturbation moves outward, away from the horizon, at least until time $t.$ At that point the complexity is minimal and beyond that it begins to increase. We may regard this as evidence  that the signal sent by Alice does not propagate to an advantageous point for signaling Bob because it is going in the wrong direction.

\begin{figure}[h!]
\begin{center}
\includegraphics[scale=.4]{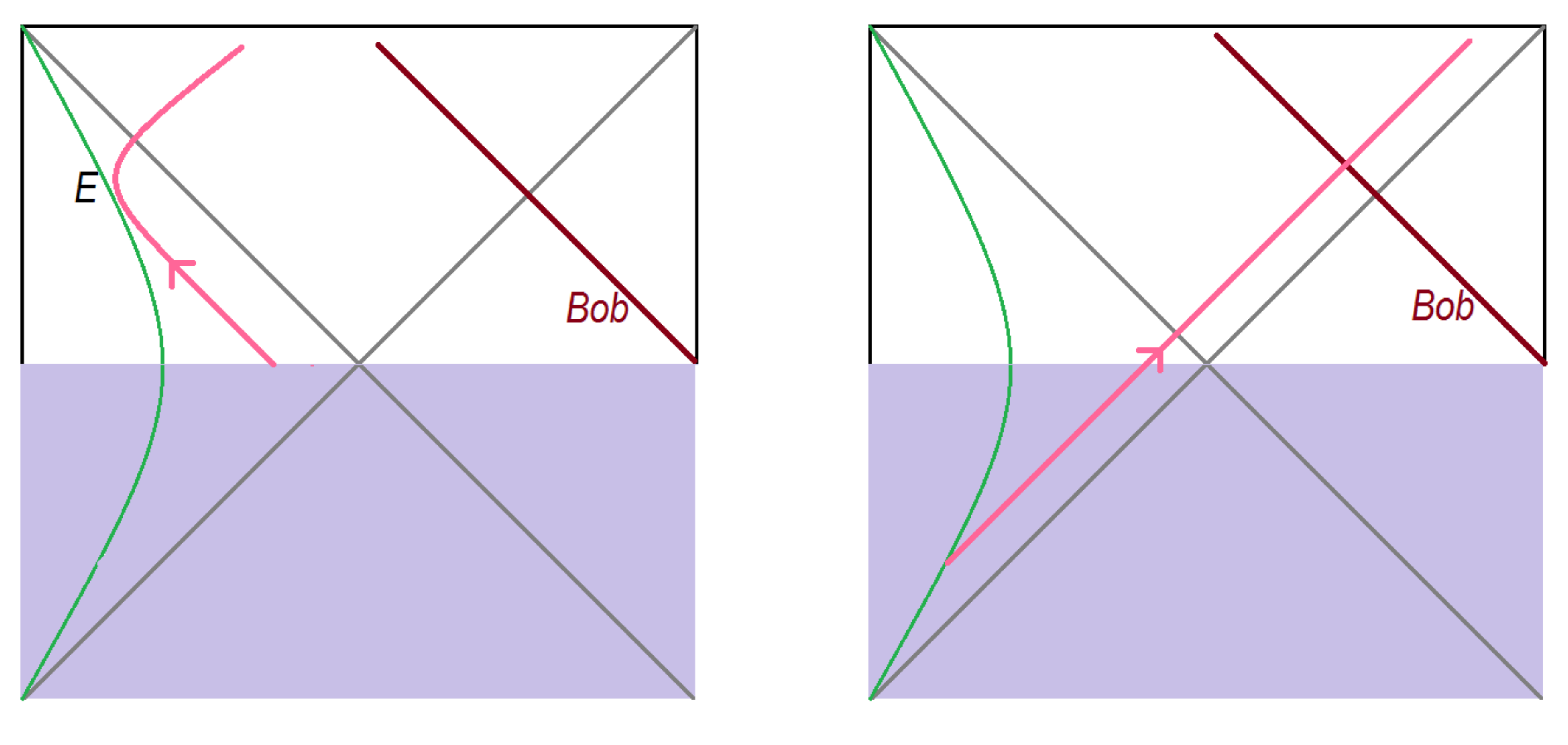}
\caption{Two kinds of perturbations on the stretched horizon shown as green. Left side:  Acting with an easy operator at $t>0$ sends a signal to the left where it cannot be received by Bob.   \ \ \ \ \
Right side: Acting with a complex future-precursor $E_P^+$ has the same effect as as acting earlier with the easy operator $E.$
Bob gets the message.   }
\label{f6}
\end{center}
\end{figure}

Note that applying an easy operator creates a state which is a local minimum of the complexity with respect to time evolution. The complexity can be assumed to increase in either direction. This is the counterpart to the statement that a particle moving from near the horizon toward the boundary of ADS will only get so far, and then turn around and fall back to the horizon. When the theory is regulated, the turn-around point will be at the boundary of the stretched horizon since the region beyond that has been excised by the coarse graining. This bounce is shown in the right side of figure \ref{f6}.


For her second attempt Alice utilizes a different strategy. Instead of acting with an easy operator at time $t$ she acts with a certain future-precursor at time $t.$
This time the precursor (which acts at positive time), is defined in terms of a fictitious easy operator, again called $E,$ but now  fictitiously acting at negative time\footnote{In \cite{Maldacena:2013xja} and \cite{Susskind:2013lpa} the operator was called $A'$ and its future precursor was called $A''.$}. Had a fictitious Alice acted on the fictitious past by applying the fictitious easy operator at time $-t,$ her message would get to Bob. Indeed if the $t\geq t_{\ast}$ the message will be a violent one \cite{Shenker:2013pqa}\cite{VanRaamsdonk:2013sza}.

In reality Alice does not act with $E$ at negative time; she acts with the future precursor $E_P^+$ at positive time,

\be
E_P^+ = U(2t) E U^{\dag}(2t).
\ee

   \bigskip
 \noindent
The operator $E_P^+ $ is constructed to have exactly the same effect as $E$ acting at $-t.$ But unlike $E$ the precursor is a very complex operator, and Alice has to be careful to get it precisely right, because the chaotic dynamics will defeat her if she makes a tiny  mistake.
 The relation between $E,$ $E_P^+,$ and Bob's trajectory are shown in figure \ref{f6}.
This time the complexity flows from simple in the past to complex in the future. This does not make it easy for Alice. Her job is to adjust $E_P^+$ so that it corresponded to something simple in the past. It's like adjusting the state of a room full of air so that if it were run backward, an over-dense pocket of air would materialize, at some specified past time, in the  corner of the room.

The evolution of complexity in this case corresponds to a signal, which at early time was propagating toward the horizon; not as in the previous case, away from the horizon. We will take that to mean that Alice has successfully launched her message.

It is now possible to state the principle that distinguishes hard and easy operators \cite{Susskind:2013lpa};  that is, operators that can send a message to Bob,  from hard operators  which cannot. We imagine the operators act at a positive physical time $t.$


\bigskip

\it
Easy operators are those which are minimally complex. If evolved in either direction their complexities would increase. In particular if formally evolved to the fictitious past their complexity would increase.

\bigskip

Hard operators  are highly complex and  become less  complex when evolved backward in time. They are fine-tuned to ``reverse-diffuse" when evolving toward the past.
\rm


\bigskip

\begin{figure}[h!]
\begin{center}
\includegraphics[scale=.5]{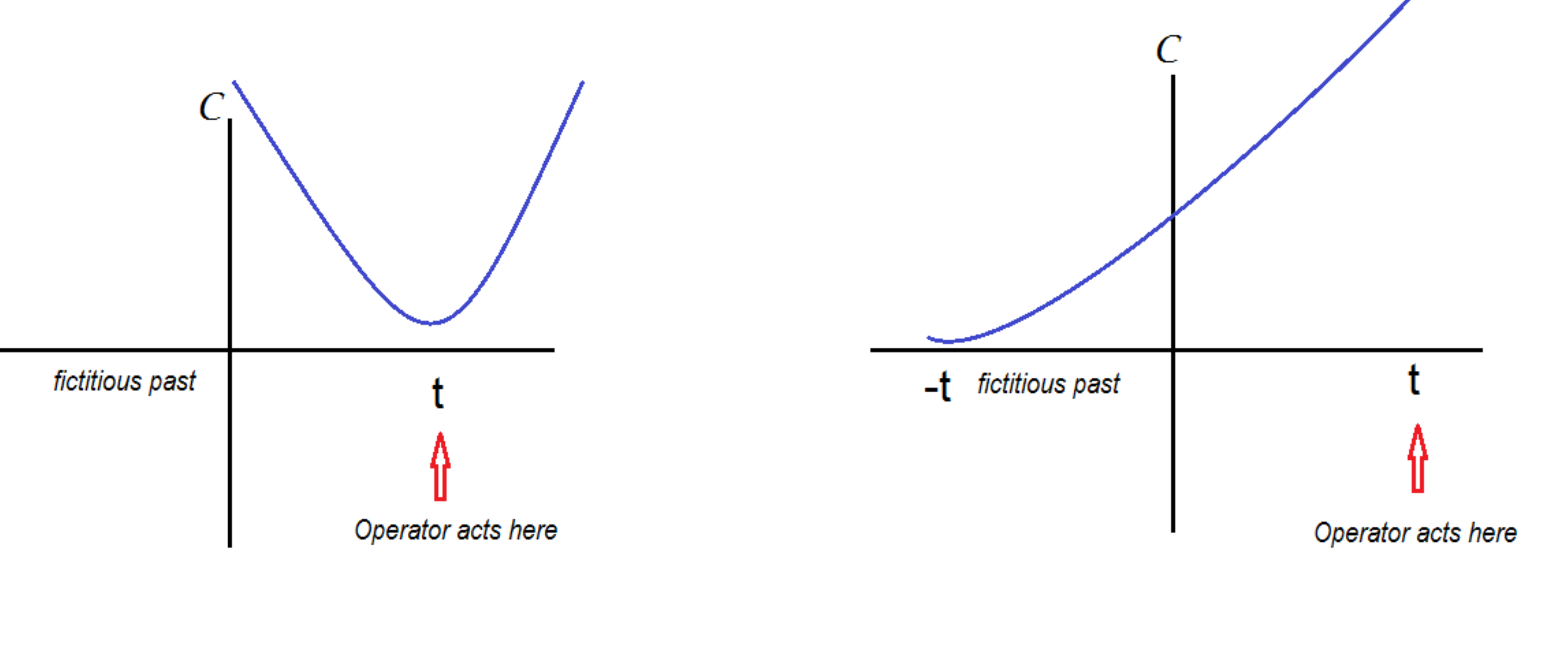}
\caption{The complexity as a function of time for two cases. In both cases an operator acts at time $t.$ In the case on the left the operator is easy and the complexity is at a minimum at $t.$  In the second case the operator is a complex precursor and the complexity decreases into the past. }
\label{C}
\end{center}
\end{figure}

The evolution of the complexity for the two cases is illustrated in figure \ref{C}.

\sc
\section{Addition to Version 2}

\bf  In \cite{Susskind:2013lpa} I pointed out that resolving the AMPSS commutator paradox \cite{Almheiri:2013hfa} requires a clear criterion for distinguishing hard and easy operators. Following \cite{Maldacena:2013xja} a possible criterion based on error correction was tentatively suggested. What was meant by an error was the interaction of a qubit with an environment. More simply, if Alice acts with an (unitary) operator made of her qubits on either the left-side black hole, or the analogous Hawking radiation of an evaporating black hole, we can think of that action as an error. Any error of this type can be corrected; all we have to do is apply the inverse unitary operator. Error correction is only interesting if one does not know whether an error has occurred.

Too much depends on somebody's state of knowledge when deciding if a certain error is correctable. Who that somebody is, and who does the error correcting, is totally unclear. For this and other reasons, error correctability  seems an unlikely  to provide a  precise criterion to distinguish whether a certain operation can send a signal through the bridge.

In this paper, at the end of  the previous section  I advocated a different criterion based on complexity and its evolution. It is interesting to compare the two criteria---error correctability and complexity evolution---to see if they give similar results.
 Depending on detailed assumptions, errors in fault tolerant codes can be corrected if they involve no more than half the qubits in the code. If one knows where the errors occured, but not what they are, then  errors involving fewer than half the qubits can be corrected. If one does not know which qubits the errors occured on, then errors involving fewer than a quarter of the qubits are correctable.  In either case their is a sharp transition and above a certain fraction  of of the total number of qubits, we expect that a perturbation will send a signal through the Einstein-Rosen bridge. In the evaporating black hole case, an interaction of the environment with more than half the photons  could  send a firewall according to criterion of error correctability.
     A possible example would be the radiation passing  through a dense dust cloudand being scattered.

 The criterion advocated in this paper  is different. It agrees with the error correction criterion for disturbances involving few qubits, but it disagrees for more massive disturbances.
  Let's consider the extreme case of acting with a product of all of the qubits. That would seem to be about as  serious disturbance as we can make. But although
   it involves every qubit, the resulting product operator is of very mild complexity. It can obviously be implemented with $N^2$ gates: one gate per qubit. Depending on whether we use the series or parallel definition  of complexity in Section 3, the complexity of the product is

   $$\c =N^2$$

   \bn
    or

    $$\c = 1.$$

   \bn
   In either case this is a very modest degree of complexity. It is
   nowhere near saturating the complexity bound of

   $$\c_{max}=e^{N^2}.$$

\bn
   Therefore one can expect that such a product operator will evolve to greater complexity, both forward in time and backward. Thus, according to our criterion the product of $N^2$ qubits is an easy operator. Alice cannot send a signal by acting with it. With this criterion interaction of the Hawking radiation with an ordinary environment will also not create a firewall at Bob's end.

It should be noted that precursors, which do send signals, typically involve all the qubits, but if $t>t_{\ast}$ they are far more complex than the simple product of qubits. They are   sums of products with extremely fine-tuned coefficients. Precursors are not likely to be encountered in nature.

   \rm

\sc
\section{Butterflies and Commutators}

Given that the system is chaotic, and that $t$ is large, the operator $E_P^+$ that Alice has to apply (in order to send a message to Bob) is not only   complex; it is  fine-tuned: the larger the time $t$ the more fine-tuned.
If  Alice makes even a tiny mistake, the backward evolution will miss the target (the easy operator $E$) by a very wide mark.  Although in principle causality does not prevent Alice from sending a message to Bob, in practice complexity makes it extremely difficult. In fact the stronger the signal Alice wishes to send \cite{Shenker:2013pqa}, the greater the fine-tuning.

What seems surprising is that Alice will be able to send the message at all by applying a very complex operator, even though no simple operator can send  a message. After all, in some sense the complex operator is just a combination of simple operators. But this  is no different than saying: There exist very fine-tuned complex operations, made of many simple operations, which can cause the air molecules in a room to clump in the corner, despite the fact that no simple operation can.

Suppose that just before Alice had applied the future-precursor $E_P^+$  an easy perturbation $e$ had acted\footnote{For simplicity the operator $e$ is assumed to be both unitary and Hermitian.}. Now consider what happens if $E_P^+$ is evolved backward in the presence of $e.$ The situation is  analogous to  attempting to run the state of a roomful of molecules back to some low entropy state. The smallest error will reverse the direction of diffusion after a short time, and undo the attempt.  Alice's message, in this case, will fail to get through, because the backward evolution will completely miss $E.$

Alice can of course correct for the error. The corrected version of the future precursor is,

\be
E_P^+(\rm corrected \it) = e \ E_P^+  \ e
\label{correction}
\ee

\bigskip
 \noindent
 instead of $E_P^+.$
 The butterfly effect implies that if $t$ is larger than the scrambling time, $E_P^+(corrected )$  will be extremely different from $E_P^+ .$ Therefore, as observed by AMPSS, the commutator

  \be
  [e, E_P^+ ]
  \label{commutator}
  \ee

   \bigskip
 \noindent
  must be large  \cite{Almheiri:2013hfa}. Indeed we may expect that the commutator of any $e$ with every $ E_P^+ $  to be large, since any small perturbation will have a large effect on every future precursor\footnote{After this paper was written I became aware of the following comment in AMPSS concerning the connection between the commutator and the butterfly effect:  "It may seem odd that measurement of a single bit can perturb many others, but this seems to be a manifestation of the butterfly effect: perturbation of a single bit, followed by a scrambling operation, perturbs all bits."}.
  
 \subsection{Addition to Version 2}
 
 \bf
  According to the interpretation of this paper, the non-vanishing commutator \ref{commutator} does not imply  $e$ creates disturbances behind Bob's horizon if it acts alone. The operator $e$ is an easy operator for which the discussion of Section 6 applies. AMPSS \cite{Almheiri:2013hfa} comes to exactly the opposite conclusion: the non-commutation of $e$ with every $E_p^+$ implies that any easy disturbance will corrupt every mode behind Bob's horizon and create a firewall \cite{Braunstein:2009my}\cite{Mathur:2009hf}\cite{Almheiri:2012rt}. The difference of interpretation can be explained as follows:
  
  AMPSS were reacting to a straw man; namely the proposal that is often called $A=R_B.$ In the two sided ADS \cite{Maldacena:2013xja} case this was expressed as $A=A''$ where $A''$ is a precursor that can send a particle to the mode $A$ on Bob's end of the Einstein Rosen bridge. The idea is that degrees of freedom behind the horizon are literally equivalent to degrees of freedom, either in the Hawking radiation ($A=R_B$) or in the stretched horizon of Alice's black hole (A=A''). This means that everywhere that the mode $A$ appears one should substitute $A''.$ Moreover the connection between $A$ and $A''$ is absolutely fixed, and not susceptible to corrections of the type in \ref{correction}. It this were correct, then indeed acting with $e$ would disturb every $A$ and create a firewall. There are several reasons to believe that this literal interpretation of $A=R_B$ is not correct including the fact that it would make Einstein-Rosen bridges traversable \cite{Maldacena:2013xja}\cite{Susskind:2013lpa}. 
  
  The interpretation of this paper is completely different. When a disturbance like $e$ occurs before $A''$ it should be absorbed into the time evolution which defines the precursor of a given operator. The form of the precursor explicitly  depends on whether $e$ acted before $A''.$ This is sometimes referred to as state dependence or non-linearity of the mapping relating the CFT  to the degrees of freedom behind the horizon. The authors of \cite{Almheiri:2013hfa} where quite aware of the issue of state dependent non-linearity  but rejected it as being inconsistent with the rules of quantum mechanics. However, some progress at formulating a state dependent map has been made by Papadodimas and Raju \cite{Papadodimas:2013jku}\cite{Papadodimas:2013wnh}.
  
  We will have to wait and see if non-linearity of the map  is viable. If, like cloning, it is something that only happens when one tries to force a global quantum description for a geometry with disconnected causal patches (observers who fall out of causal contact), but is never observed within any single causal patch, then I see no reason to reject it.
  \rm

\sc
\section{Reverse-Diffusion Interrupted}

Since the laws of physics are reversible anything that can happen, can happen in reverse\footnote{This section summarizes one of the findings of Shenker and Stanford in the language of complexity and chaos. It is explained much more completely in \cite{SS}}. For example consider a tub of water in thermal equilibrium. At time zero, drop a bit of black ink into it. The ink will diffuse, with the density given by

\be
\rho = t^{-3/2} e^{-r^2/t}
\label{diff}
\ee

\bn
Eventually the blob will be completely thermalized.

It follows that the reverse diffusion process described by

\be
\rho =(T- t)^{-3/2} e^{-r^2/(T-t)}
\label{revdiff}
\ee

\bn
is also possible. Equation \ref{revdiff} describes a fine tuned initial starting point in which a diffuse cloud of ink assembles itself into a shrinking blob, which shrinks to a point at $t= T.$ Call it a reverse-diffuser or $RD.$
This reverse diffusion is extraordinarily difficult to set up, but it is every bit as good a solution as the forward diffusion process.

What happens if a delicately adjusted RD runs into an ordinary  forward-diffusing blob  (FD); who wins?  The answer depends on the  timing of the two blobs. If FD blob is introduced locally, shortly  before the reverse blob shrinks to a point, the RD may win and continue to shrink, although it may be delayed.

On the other hand if the FD is introduced while the RD is still extremely mixed throughout  the water,  the delicate initial conditions will be upset, and in a short time it will join the forward-diffusing blob. The butterfly effect will have done its work and prevented the reverse diffusion from completing its mission (of creating a point-like disturbance).

\begin{figure}[h!]
\begin{center}
\includegraphics[scale=.3]{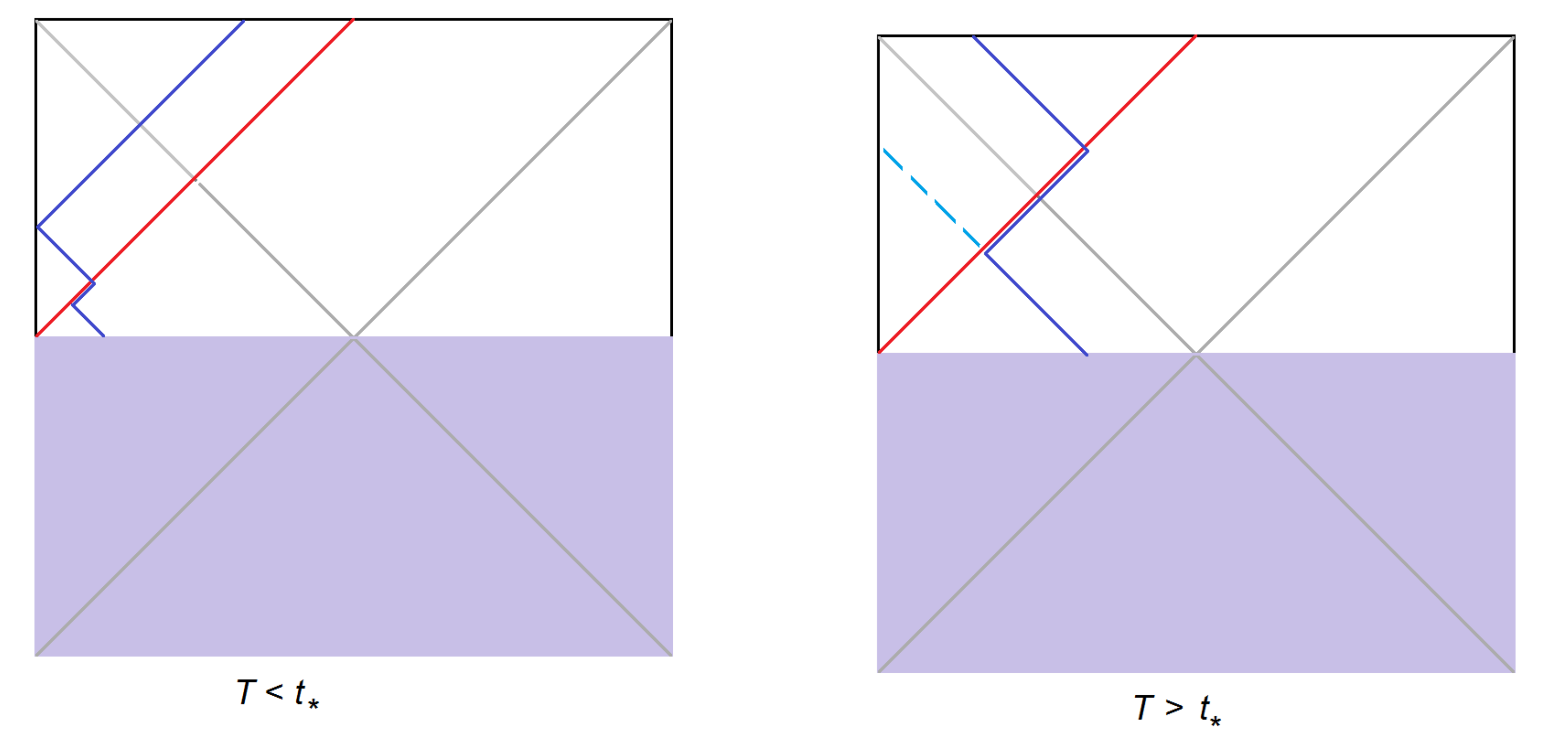}
\caption{ }
\label{interrupt}
\end{center}
\end{figure}

Now we turn to the bulk gravity description of ADS black holes.
Consider two bulk gravitational  situations. In the first, a simple   perturbation has been launched at $t=0$ from near the UV boundary \cite{Shenker:2013pqa}\cite{VanRaamsdonk:2013sza}.  It moves toward the horizon, paralleling the evolution of the forward-diffusing blob. Eventually it is lost into the horizon. It has behaved exactly as the FD.

Next consider the history created by a precursor acting at $t=0. $ The precursor is constructed
 so as to become a simple perturbation at $t=T,$

 \be
E_P^- = U^{\dag}(T) E U(T).
\label{E'=UEU^*}
\ee

 \bn
 From the bulk point of view we can think of the precursor as launched (at t=0) from a region  close to the horizon, and moving toward the boundary. It is tuned to  get to the boundary at $t=T$ and then turn around. The similarity with the reverse-diffusing RD is obvious, including the bounce at time $T.$

Finally, consider what happens if both perturbations are introduced into the initial state at $t=0$.  The corresponding bulk gravity problem of evolving the combined disturbance  has been solved by Shenker and Stanford \cite{SS}. I will merely indicate their result with the two diagrams in figure \ref{interrupt}:

 If we take  $T< t_{\ast}$ then after a delay the  precursor will make it to the boundary. The delay is small if $T<< t_{\ast}.$ The delay grows with $T,$ and when  $T$ reaches the scrambling time $ t_{\ast}$, the second pulse becomes completely reversed before reaching the boundary. Instead  it is swept along with the first pulse,  to the singularity.   This behavior is just what we would expect from the butterfly effect.

 Stated in terms of complexity, if $T > t_{\ast}$ the complexity of the precursor never gets much smaller than \ref{Cscram}. In terms of water and ink, the water remains cloudy throughout the process.

\sc
\section{A Complexity Horizon}
\begin{figure}[h!]
\begin{center}
\includegraphics[scale=.3]{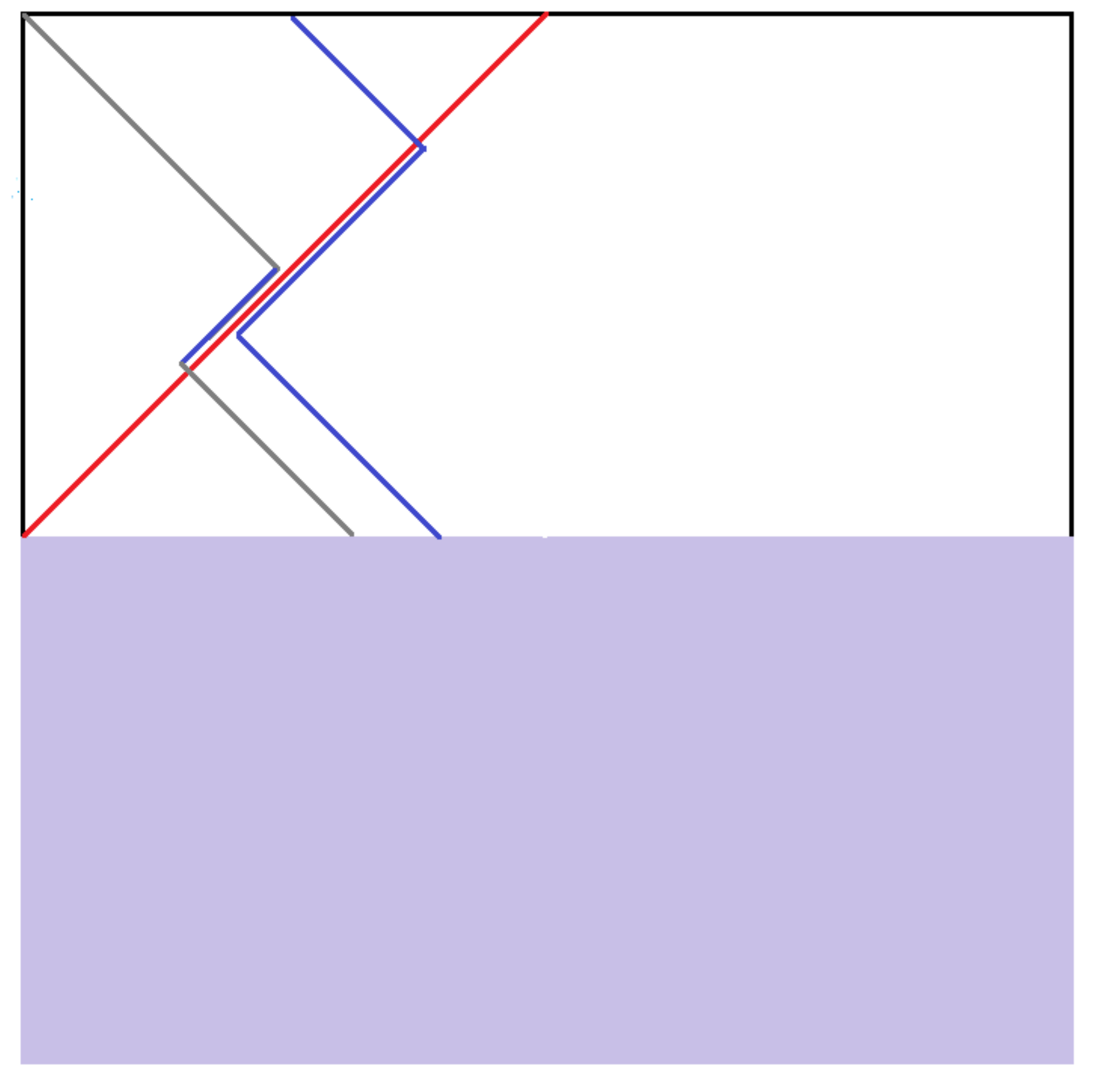}
\caption{ }
\label{interrupt2}
\end{center}
\end{figure}

The Shenker-Stanford effect described above gives a new perspective of the meaning of a horizon. The diagram on the right side of figure has not accounted for the back reaction on the geometry. The diagram is not a properly drawn Penrose diagram for the obvious reason that the blue light ray has not been drawn as a straight line. One can either correct the Penrose diagram or work with the diagram as drawn, recalling that whenever a trajectory crosses the red shockwave, it has to be displaced by the appropriate amount.
This includes the light rays that define the horizon of Alice's black hole. In figure \ref{interrupt2}
the figure is redrawn to include the horizon shift. The interesting point, noted by \cite{SS},  is that for $T>t_{\ast}$ the precursor disturbance never appears outside the horizon of Alice's black hole.
This is to be compared with the discussion of water and ink. If there is enough time for the butterfly effect to act,  the FD wave will turn the RD wave around, so that it does not have a chance to become localized.

One might say that there is a complexity horizon. Disturbances which can assemble into simple configurations are on the near side of the complexity horizon, and are eventually visible to crude observations. However the complexity horizon can be shifted by a forward diffusing perturbation so that the original perturbation is pushed back behind the complexity horizon.


\sc
\section{Very Large Time}

Complexity reaches an upper limit at time $\omega = e^S.$ One can see this a number of ways. One way is from the fact that any unitary operator can be constructed from $e^S$ gates, but there are other ways to see that something happens at this time. First of all $t_{cr}=e^S$ defines the classical recurrence time. It is not entirely clear what this has to do with quantum complexity. I will describe a more relevant effect. Consider  a chaotic system with $e^S$ non-degenerate energy levels $E_i.$ The average separation between levels is $\delta E \sim e^{-S}.$

Let the state at time $t=0$ be,

\be
|\Psi (0)\ra = \sum_i F_i |i\ra
\ee

   \bigskip
 \noindent
Assume that the  $F_i$ are all real.
The time evolution in this basis is very simple; each term picks up a phase,

\be
|\Psi (t)\ra = \sum_i F_i e^{-iE_i t} |i\ra
\ee

   \bigskip
 \noindent
The entire evolution is reduced to the motion of a point on an $e^S$ dimensional torus, with velocities given by

\be
v_i =  E_i.
\label{velocity}
\ee

\begin{figure}[h!]
\begin{center}
\includegraphics[scale=.3]{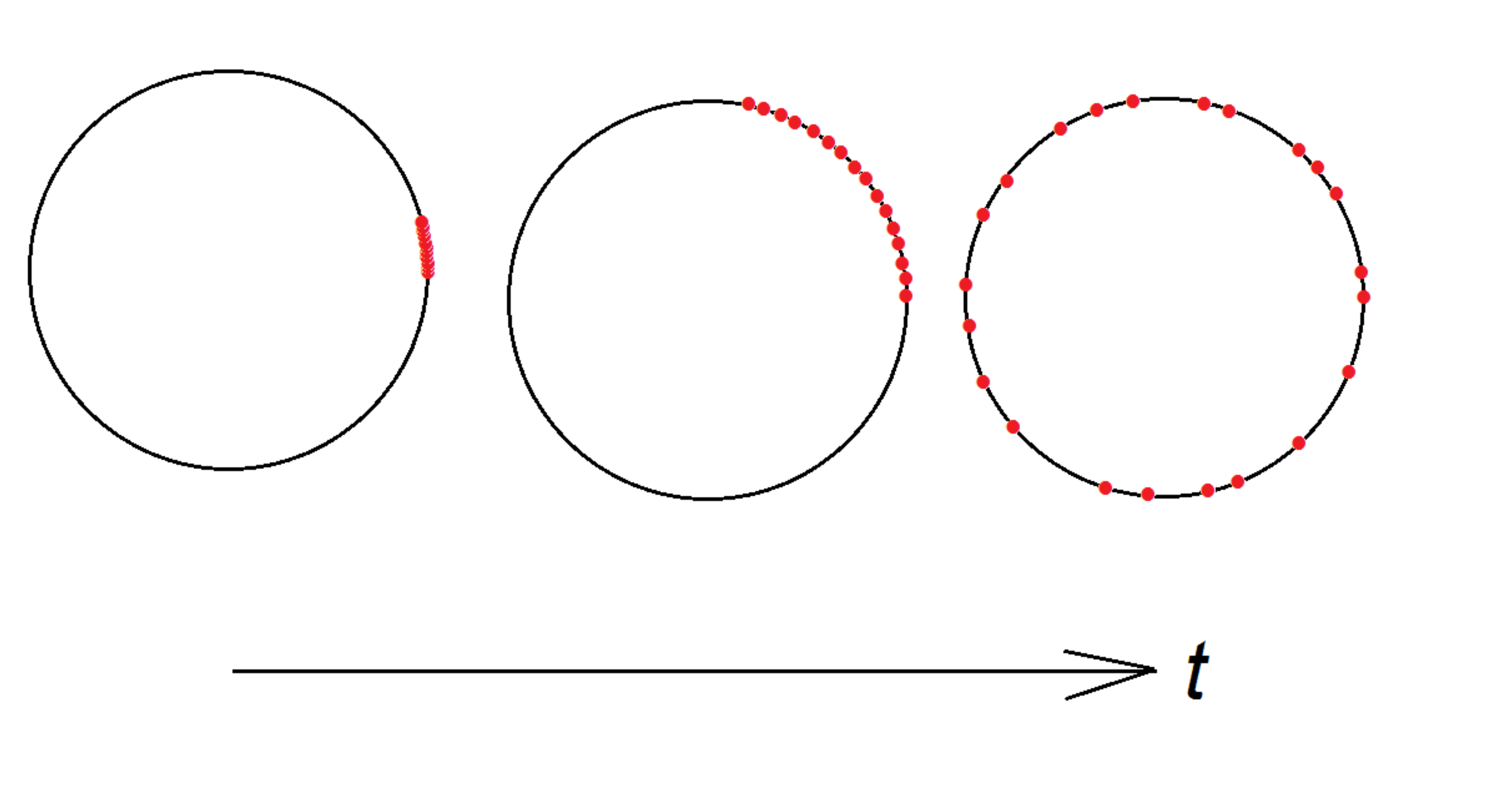}
\caption{Evolution of phases }
\label{f7}
\end{center}
\end{figure}

   \bigskip
 \noindent
We can also picture it as $e^S$ points moving on a circle as in figure \ref{f7}, with   velocities \ref{velocity}. The energy levels are incommensurate so that the motion is ergodic. At first it is very orderly since neighboring points have extremely small relative velocities. During this time the complexity increases.
For a long time neighbors remain close, but at a time of order the classical recurrence time

\be
t_{cr} \sim e^S
\ee

   \bigskip
 \noindent
the neighbors will become separated by angles of order $\pi.$ Beyond that time the orderly organization of the phases is no longer recognizable. The configuration looks like a random sprinkle of points on the circle. During this time there are small fluctuations of the complexity but with no pattern of increase or decrease.

Let us consider what this means for Alice. If she waits a time, of order $e^S,$ before applying the precursor,  there will be no overall pattern to the time dependence of the complexity of $E^+$. There will be extremely small fluctuations, but no net tendency to either increase or decrease.   It is not completely clear what this means for Bob and I won't try to speculate.

On even longer time scales the phases undergo quantum recurrence. The quantum recurrence time $t_{qr}$ is doubly exponential,

$$ t_{qr} = e^{e^S}.$$

   \bigskip
 \noindent
On this time scale the phases will intermittently return to a configuration in which they are all close to being real. In other words $U(t)$ will return to unity and the complexity will become small. During this period Bob should be bombarded with signals. But for even larger $t$ the complexity will turn around so that Bob will see nothing; and so on, to infinity \footnote{I am grateful to Douglas Stanford for explaining this and other related points to me.}.

.

\sc
\section{Conclusion (updated  in V2) }

 A firewall is a blast of particles propagating almost parallel to the horizon, but just behind it.
In \cite{Maldacena:2013xja} it was argued that the only mechanism for creating such particles is to have them originate at the opposite end of an Einstein-Rosen bridge. This paper argues that these effects are symptoms of one of two extreme situations:

\bn
1) Extreme fine-tuning of initial conditions, of the kind that lead to apparent local violations of the second law such as reverse-diffusion and white holes;

\bn
or

\bn
2) Exponentially long time intervals that lead to Poincare recurrences and freak phenomena such as Boltzmann brains. Given enough time the freak phenomena become generic. The same may be true for firewalls.

\bn
The first of these conditions, although not impossible,  is normally disallowed, being too unlikely to ever occur. The second may be relevant for eternal black holes, but not for evaporating black holes, which disappear long before the recurrence time.

I have not discussed in any detail the evaporating black hole case. The
biggest obstacle  is the lack of understanding  of the  Einstein-Rosen bridge between a black hole to its  cloud of Hawking radiation. Such a bridge is implied by the entanglement of the two \cite{Maldacena:2013xja}. Hopefully this will be remedied. However there is reason to believe that the transfer of qubits from the black hole to the radiation makes it, if anything, more difficult to send a message.

\section*{Acknowledgements}

I thank Douglas Stanford and Steve Shenker for many discussions about the butterfly effect. Some of the material throughout the paper is based on extensive conversations with them, and their work.

Support for this research came through NSF grant Phy-1316699 and the Stanford Institute for Theoretical Physics.

\end{document}